\documentclass{appolb}
\usepackage{graphicx}

\def\ra{\!\rightarrow\!}

\def\simge{\mathrel{%
   \rlap{\raise 0.511ex \hbox{$>$}}{\lower 0.511ex \hbox{$\sim$}}}}
\def\simle{\mathrel{
   \rlap{\raise 0.511ex \hbox{$<$}}{\lower 0.511ex \hbox{$\sim$}}}}

\begin{document}


\title{
\vskip-2.5in
\begin{flushright}
\normalsize{\rm
UCHEP--19--01 \\
}
\end{flushright}
\vskip0.40in
Semileptonic and leptonic charm meson decays at Belle~II\thanks{Presented at 
the Tenth International Workshop on the CKM Unitarity Triangle, September 17-21, 2018, 
Heidelberg, Germany.}
}

\author{A.~J.~Schwartz
\address{Physics Department, University of Cincinnati, 
Cincinnati, Ohio 45221 USA}
}

\maketitle

\begin{abstract}
We review measurements of semileptonic and leptonic charm meson 
decays performed by the Belle experiment, and we use these results
to estimate the sensitivity of the follow-on Belle~II experiment 
to these decays.
\end{abstract}


\section{Introduction}

Semileptonic and leptonic $D$ meson decays are easier to understand 
theoretically than hadronic decays, as the hadronic uncertainties 
factorize. They are also straightforward 
to measure at an $e^+e^-$ experiment due to low backgrounds and 
good detector hermeticity. They have been studied at experiments
CLEOc~\cite{cleoc}, BESIII~\cite{bes3}, Belle~\cite{belle}, 
and Babar~\cite{babar}, and they constitute 
an important part of the physics program of Belle~II~\cite{belle2}. 
The Belle~II experiment runs at the SuperKEKB accelerator at the KEK 
laboratory in Japan and is the follow-on experiment to Belle.
The accelerator collides 4~GeV/$c$ positrons with 7~GeV/$c$ 
electrons; the center-of-mass energy is tuned to be at the $\Upsilon(4S)$ 
resonance in order to produce copious amounts of $B$ mesons
via $e^+e^-\ra\Upsilon(4S)\ra B\bar{B}$. 
The Belle~II detector is now being commissioned and will begin
taking physics data in the spring of 2019. In this paper we review 
measurements of leptonic and semileptonic charm decays made 
by the preceding Belle experiment, and we use these 
results to estimate the expected sensitivity of Belle~II.

\section{Leptonic decays}

The partial width $\Gamma(D^+_{(s)}\rightarrow\ell^+\nu)$~\cite{charge-conjugates} 
is given by the formula~\cite{theory:lepton_formula}
\begin{eqnarray}
\Gamma(D^+_{(s)}\ra\ell^+\nu) & = & \frac{G^2_F}{8\pi} 
f^2_{D_{(s)}} |V^{}_{cx}|^2 m^{}_{D_{(s)}}\,m^2_\ell 
\left(1-\frac{m^2_\ell}{\ \ m^2_{D_{(s)}}}\right)^2 ,
\label{eqn:leptonic} 
\end{eqnarray}
where $f^{}_{D_{(s)}}$ is the $D^+_{(s)}$ decay constant,
and $V^{}_{cx}$ is the Cabibbo-Kobayashi-Maskawa (CKM) 
matrix element $V^{}_{cs}$ for $D^+_s$ decays and 
$V^{}_{cd}$ for $D^+$ decays~\cite{pdg:ckm}. The decay 
constant $f^{}_{D_{(s)}}$ parameterizes the hadronic 
matrix element $\langle 0|{\cal H}|D^+_{(s)}\rangle$.
To test the Standard Model (SM), one measures the 
branching fraction $B(D^+_{(s)}\rightarrow\ell^+\nu)$, calculates the
partial width $\Gamma = B/\tau^{}_D$, and uses Eq.~(\ref{eqn:leptonic})
to determine the product $f^{}_{D_{(s)}} |V^{}_{cx}|$. One then either
takes $|V^{}_{cx}|$ from other measurements and CKM unitarity to extract 
$f^{}_{D_{(s)}}$, or takes $f^{}_{D_{(s)}}$ from lattice QCD theory to 
extract~$|V^{}_{cx}|$. 

\begin{figure}[htb]
\begin{center}
\hbox{
\includegraphics[width=6.3cm]{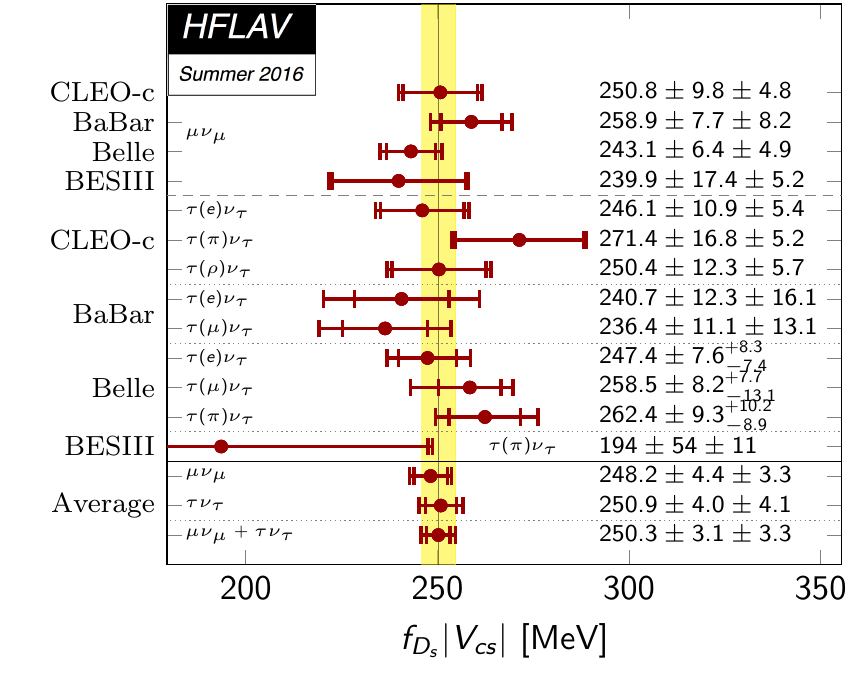}
\includegraphics[width=6.3cm]{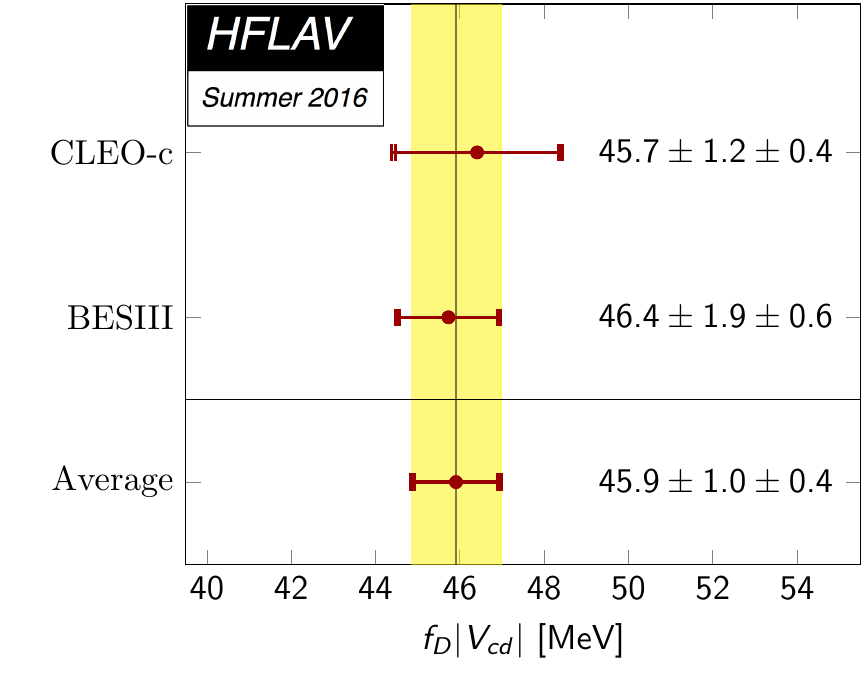} }
\end{center}
\vskip-0.30in
\caption{World average (WA) values as calculated by the 
Heavy Flavor Averaging Group (HFLAV)~\cite{hflav:leptonic} for the products
$f^{}_{D_s}|V^{}_{cs}|$ (left) and $f^{}_{D}|V^{}_{cd}|$ (right).}
\label{fig:fdotV_hflav}
\end{figure}

The Heavy Flavor Averaging Group (HFLAV)~\cite{hflav:leptonic} has calculated 
world average (WA) values of the product $f^{}_{D_{(s)}} |V^{}_{cx}|$ using 
all relevant experimental measurements; the results are shown in 
Fig.~\ref{fig:fdotV_hflav}.
The WA values are
$f^{}_{D_s} |V^{}_{cs}|= (250.3\pm 3.1\pm 3.3)$~MeV and 
$f^{}_{D} |V^{}_{cd}|= (45.9\pm 1.0\pm 0.4)$~MeV.
The Flavor Lattice Averaging Group~\cite{lattice:flag17} quotes 
$f^{}_{D_s} = (248.83\pm 1.27)$~MeV and 
$f^{}_{D} = (212.15\pm 1.45)$~MeV
based on lattice QCD results from Refs.~\cite{lattice:fnalmilc14,lattice:etm15}. 
Inserting these values gives
\begin{eqnarray}
|V^{}_{cs}| & = & 1.006\pm 0.018\pm 0.005 \\
|V^{}_{cd}| & = & 0.2164\pm 0.0050\pm 0.0015\,,
\end{eqnarray}
where the first error is experimental and the second is from theory. 
Alternatively, inserting more recent (and precise) lattice QCD results 
$f^{}_{D_s} = (249.9\pm 0.4)$~MeV and 
$f^{}_{D} = (212.7\pm 0.6)$~MeV
from the Fermilab/MILC Collaboration~\cite{lattice:fnalmilc18}
gives essentially identical results for $|V^{}_{cs}|$ and $|V^{}_{cd}|$.

Conversely, inserting CKM matrix elements
$|V^{}_{cs}|= 0.973394\,^{+0.000074}_{-0.000096}$ and 
$|V^{}_{cd}|= 0.22529\,^{+0.00041}_{-0.00032}$ 
as obtained by the CKM Fitter group~\cite{ckmfitter} from a global 
fit to various measurements subject to CKM unitarity~\cite{pdg:ckm}, we obtain
\begin{eqnarray}
f^{}_{D_s} & = & (257.1\pm 4.6)~{\rm MeV} \\
f^{}_{D} & = & (203.7\pm 4.9)~{\rm MeV}\,.
\end{eqnarray}
These values are consistent with those calculated from lattice QCD.

\begin{center}
------------------
\end{center}

\begin{figure}[htb]
\vskip-1.0in
\begin{center}
\hbox{\hskip-0.15in
\includegraphics[width=6.3cm]{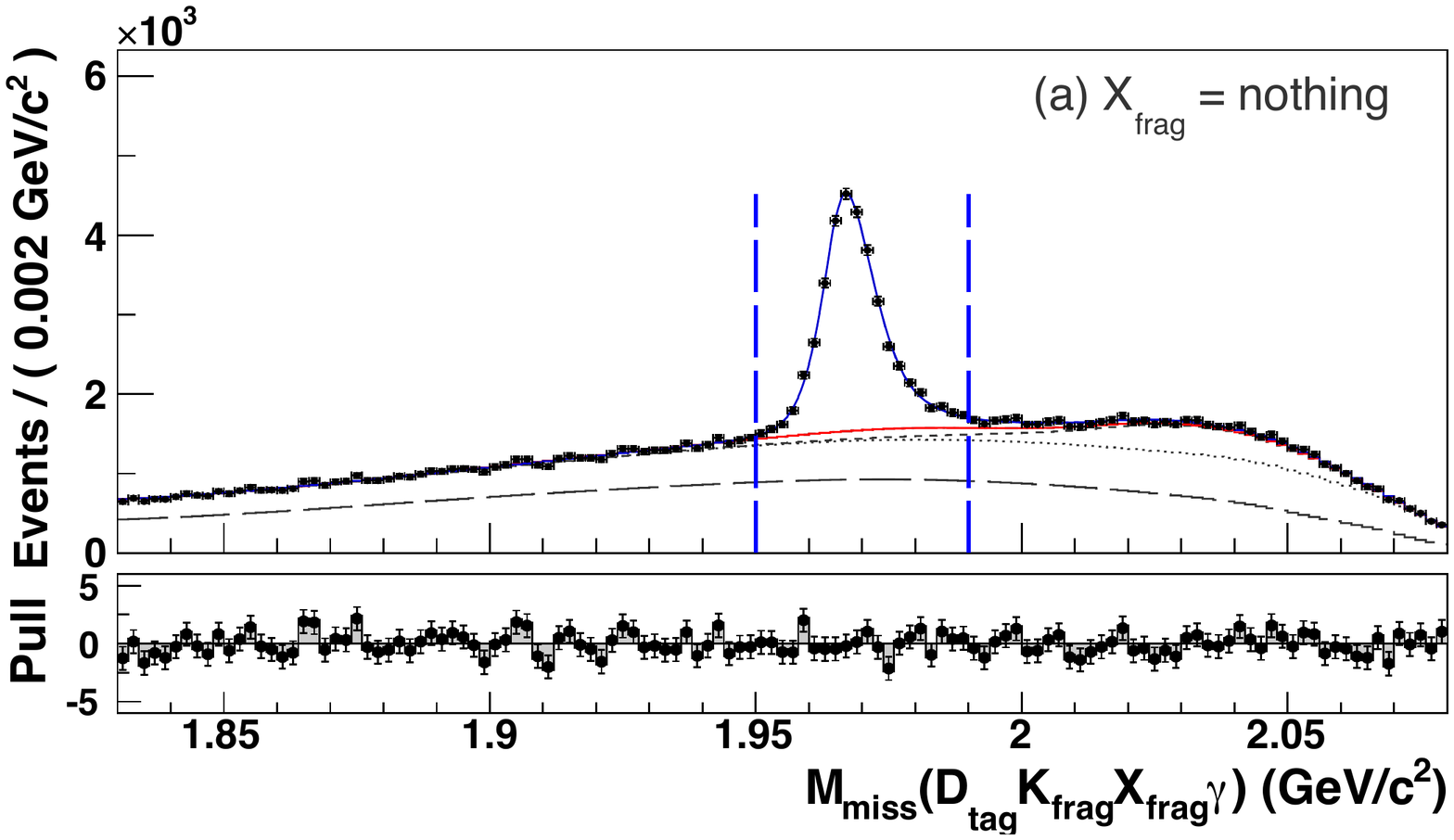}
\hskip0.10in
\includegraphics[width=6.3cm]{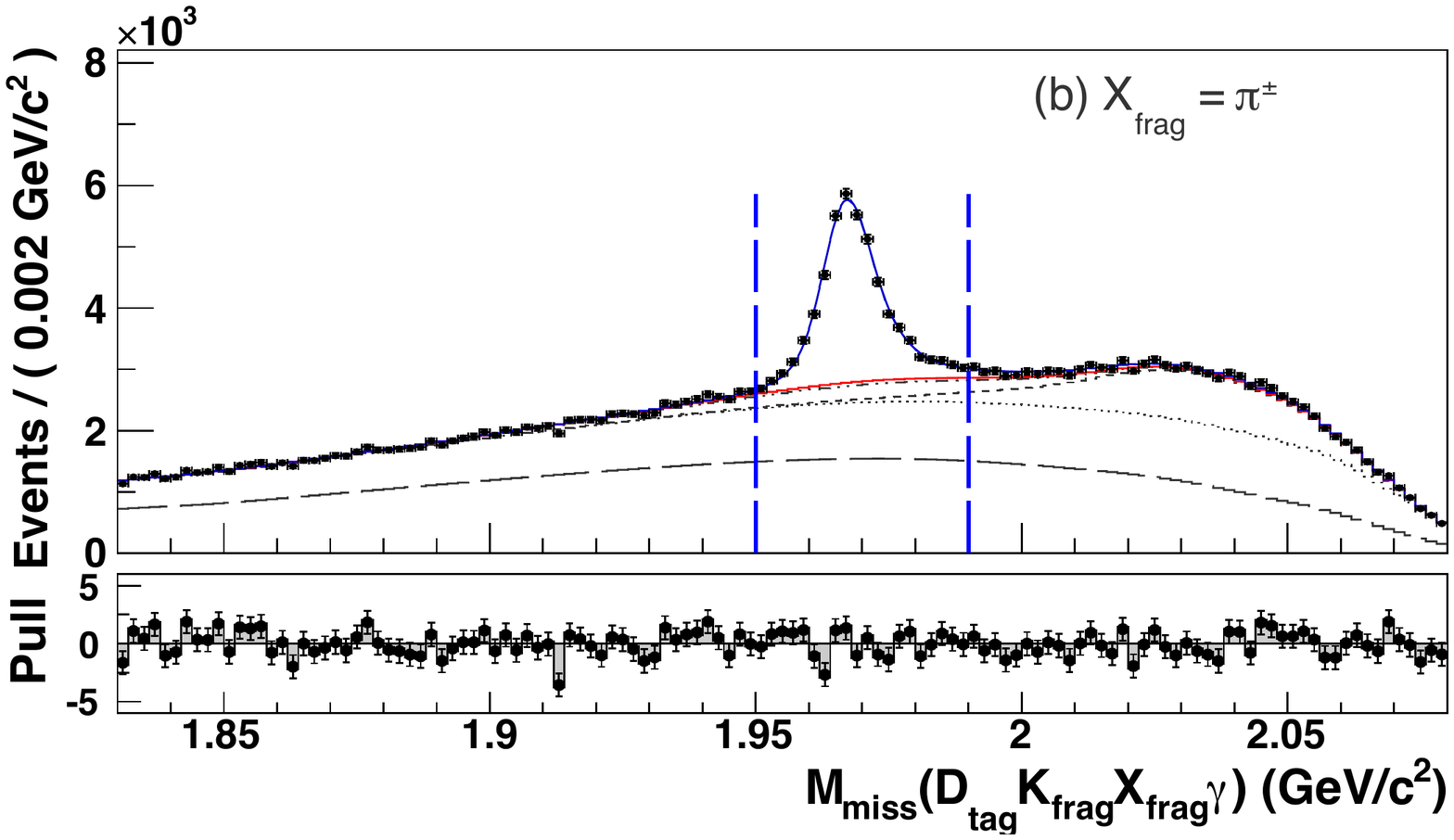}}
\vskip-1.80in
\hbox{\hskip-0.15in
\includegraphics[width=6.3cm]{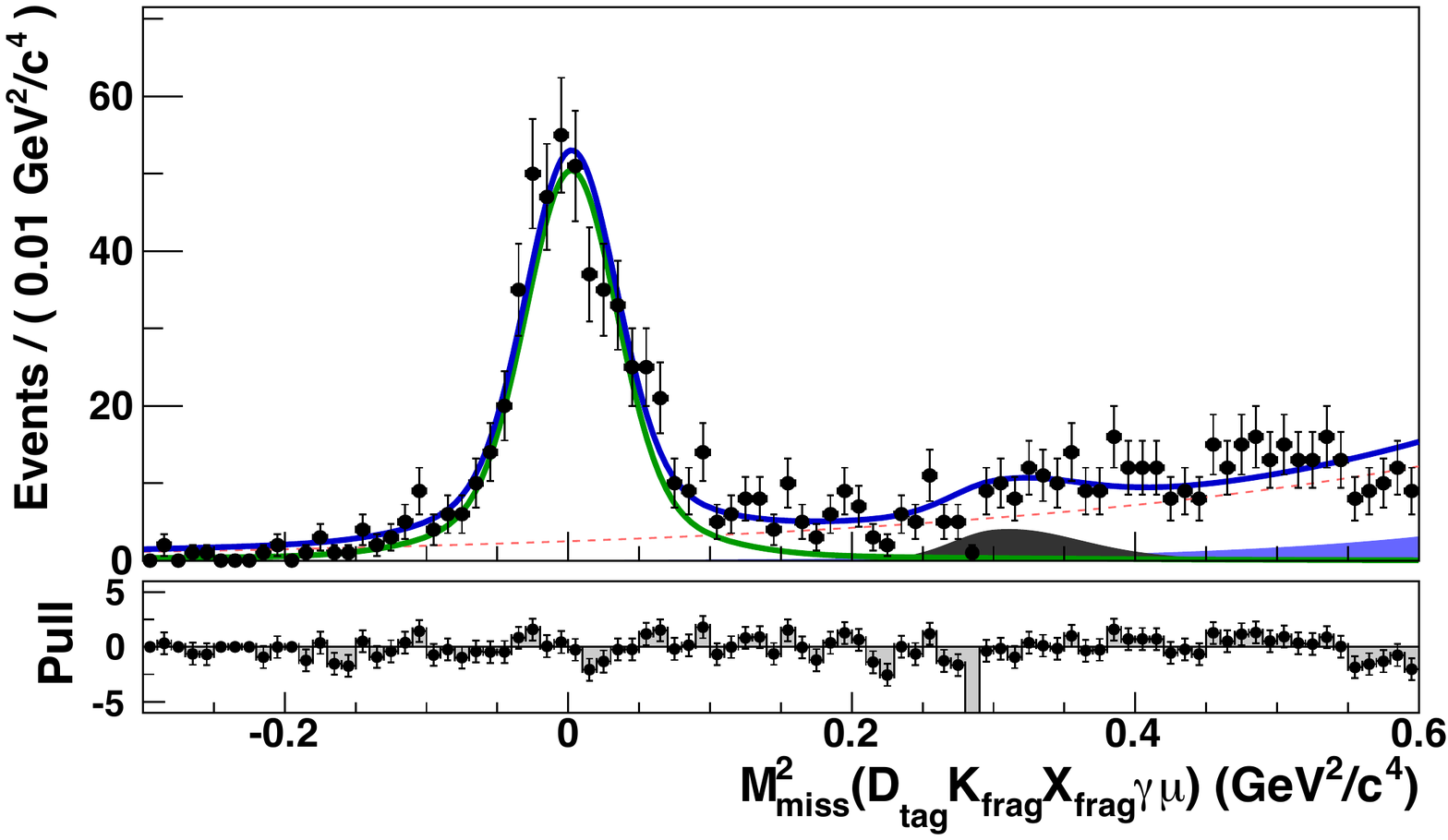}
\hskip-1.05in
\vbox{
\includegraphics[width=7.0cm]{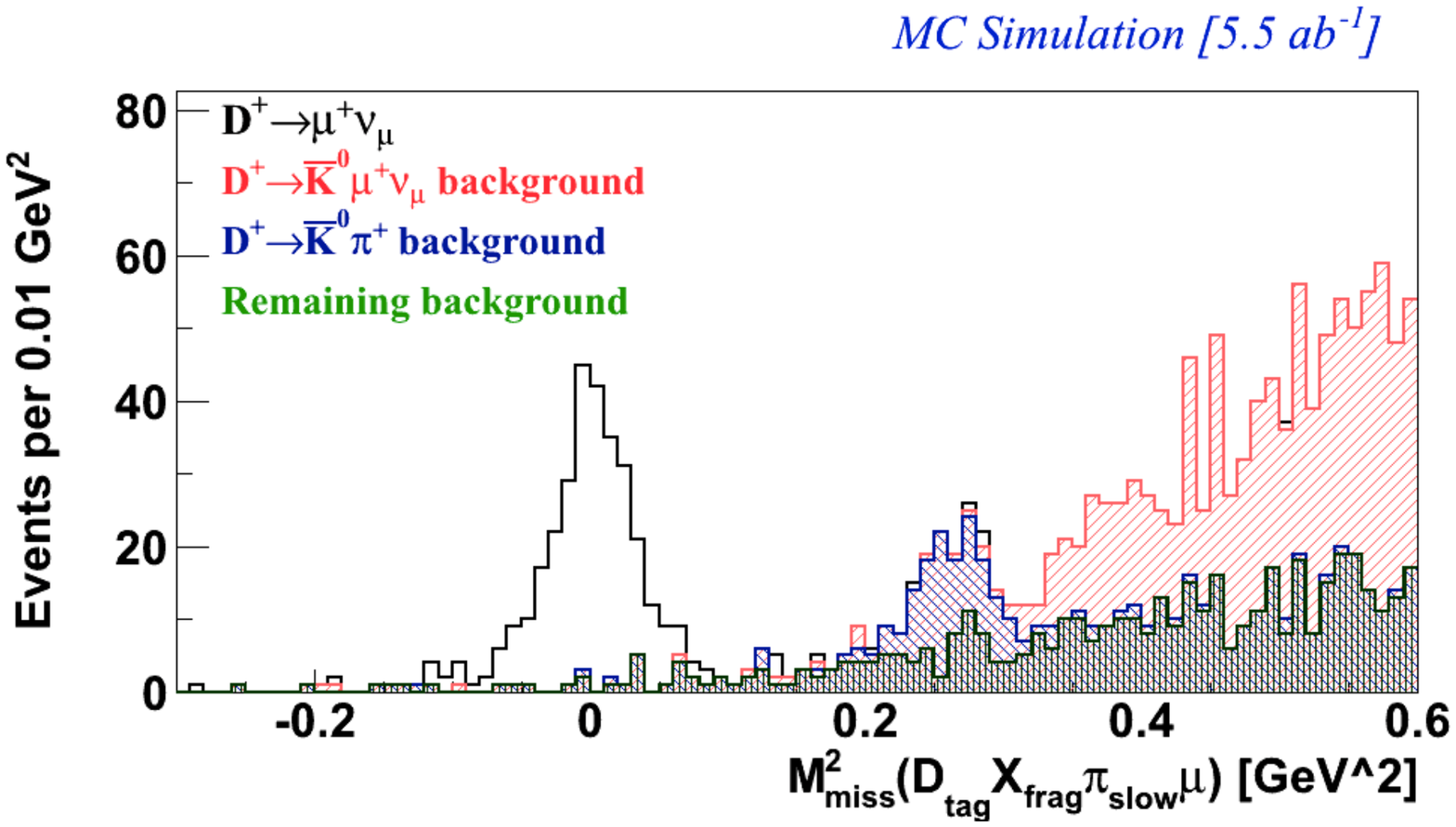} }}
\end{center}
\vskip-0.90in
\caption{Top: missing mass 
$\sqrt{(P^{}_{e^+} + P^{}_{e^-} - P^{}_{\rm tag} - P^{}_{X_{\rm frag}} - P^{}_\gamma)^2}$
from a Belle analysis of $D^{*+}_s\ra D^+_s\gamma$, $D^+_s\ra\mu^+\nu$ decays 
using 913~fb$^{-1}$ of data for two fragmentation systems 
$X^{}_{\rm frag}\!=\!{\rm nothing}$ (left) and 
$X^{}_{\rm frag}\!=\!\pi^\pm$ (right)~\cite{belle:leptonic}.
Bottom left: missing mass
$\sqrt{(P^{}_{e^+} + P^{}_{e^-} - P^{}_{\rm tag} - P^{}_{X_{\rm frag}} - P^{}_\gamma - P^{}_\mu)^2}$
from the same Belle analysis. The peak at zero corresponds to the undetected neutrino
from $D^+_s\ra\mu^+\nu$. 
Bottom right: missing mass
$\sqrt{(P^{}_{e^+} + P^{}_{e^-} - P^{}_{\rm tag} - P^{}_{X_{\rm frag}} - 
P^{}_{\pi^0} - P^{}_\mu)^2}$
from a Belle II MC study of $D^{*+}\ra D^+\pi^0,\,D^+\ra\mu^+\nu$ decays 
corresponding to 5.5~ab$^{-1}$ of data~\cite{belle2:ptep}. A significant 
signal peak is visible.}
\label{fig:belle_Dsmunu}
\end{figure}

Belle has measured 
$D^{*+}_s\ra D^+_s\gamma$, $D^+_s\ra\mu^+\nu$ decays using 913~fb$^{-1}$ 
of data~\cite{belle:leptonic}. The analysis proceeds in four steps:
\begin{enumerate}
\item a $D^{(*)0}$, $D^{(*)+}$, or $\Lambda^+_c$ decay is reconstructed on 
the ``tag-side'' of an event, i.e., recoiling against the signal-side 
$D^+_s\ra\mu^+\nu$ decay. To conserve strangeness, a $K^\pm$ or $K^0_S$ 
is also required on the tag side. If a $\Lambda^+_c$ decay were 
reconstructed, then a $\bar{p}$ is required to conserve baryon number.
\item a ``fragmentation system'' ($X^{}_{\rm frag}$) is constructed 
from 1-3 $\pi^\pm$ tracks and 0-1 $\pi^0$ candidates. From the measured 
four-momenta $P$, a ``missing mass'' $\sqrt{P^2_{\rm miss}} = 
\sqrt{(P^{}_{e^+} + P^{}_{e^-} - P^{}_{\rm tag} - P^{}_K - P^{}_{X_{\rm frag}})^2}$ is
calculated and required to be within $3\sigma$ in resolution of $M(D^{*+}_s)$.
\item a low-momentum $\gamma$ is required, presumably originating 
from $D_s^{*+}\ra D^+_s\gamma$, and the missing mass 
$\sqrt{(P^{}_{e^+} + P^{}_{e^-} - P^{}_{\rm tag} - P^{}_K - P^{}_{X_{\rm frag}}-P^{}_\gamma)^2}$
is calculated. This distribution should peak near $M(D^+_s)$ for signal decays, 
and it is fitted to obtain an inclusive $D^+_s$ signal yield. 
\item a high mometum $\mu^+$ pointing to the 
interaction point is required, and the missing mass
$\sqrt{(P^{}_{e^+} + P^{}_{e^-} - P^{}_{\rm tag} - P^{}_K - P^{}_{X_{\rm frag}}-P^{}_\gamma-P^{}_\mu)^2}$ 
is calculated. This distribution should peak at $m^{}_\nu\approx 0$ for signal decays;
it is fitted to obtain the exclusive $D^+_s\ra\mu^+\nu$ signal yield.
\end{enumerate}
The results of the third step are shown in Figs.~\ref{fig:belle_Dsmunu}a,b for 
the two simplest $X^{}_{\rm frag}$ systems. Fitting these distributions (and 
also those of the other $X^{}_{\rm frag}$ systems) yields 
$94360\pm 1310\,{\rm (stat.)}\pm 1450\,{\rm (syst.)}$ inclusive $D^+_s$ 
decays. The result of the last step is shown in Fig.~\ref{fig:belle_Dsmunu}c;
fitting this distribution yields $492\pm 26$ $D^+_s\ra\mu^+\nu$ decays.

This method can also be used at Belle~II. As the Belle measurement 
is limited by statistics rather than systematics, we scale the event 
yields obtained by Belle 
by the ratio of luminosities. The result is $5.2\times 10^6$ inclusive 
$D^+_s$ decays, and 26900 exclusive $D^+_s\ra\mu^+\nu$ decays, in 50~ab$^{-1}$ 
of Belle~II data. The latter sample should yield statistical errors 
of $\delta |V^{}_{cs}| = 0.003$ and $\delta f^{}_{D_s} = 0.8$~MeV, which 
are similar to the current theoretical errors arising from lattice QCD. 

A similar analysis was performed at Belle for $D^+_s\ra\tau^+\nu$ 
decays~\cite{belle:leptonic}. In this case a yield of $2217\pm 83$ 
exclusive decays were obtained. Scaling this yield by the ratio of Belle 
and Belle~II luminosities yields 121400 $D^+_s\ra\tau^+\nu$ decays in 
50~ab$^{-1}$ of Belle~II data. This sample size should give errors of 
$\delta |V^{}_{cs}| = 0.0014$ and $\delta f^{}_{D_s} = 0.4$~MeV, which
are twice as precise as the corresponding measurements from $D_s^+\ra\mu^+\nu$. 

For $D^+\ra\mu^+\nu$ decays, Belle did not collect enough data to 
observe this mode. For Belle~II, a Monte Carlo (MC) simulation 
study~\cite{belle2:ptep} indicates that 1250 exclusive $D^+\ra\mu^+\nu$ 
decays would be reconstructed in 50~ab$^{-1}$ of data. The corresponding
missing mass distribution is shown in Fig.~\ref{fig:belle_Dsmunu}d. 
This signal yield should result in a statistical error 
$\delta (f^{}_D\!\cdot\!|V^{}_{cd}|) = 0.68$~MeV, which is well
below the current errors from CLEOc (2.0~MeV)~\cite{cleoc:leptonic} 
and BESIII (1.2~MeV)~\cite{bes3:leptonic}.

\section{Semileptonic decays}

For semileptonic decays $D\ra K\ell^+\nu$ and $D\ra \pi\ell^+\nu$, 
the differential partial width to lowest order in 
$m^2_\ell$ is~\cite{theory:semileptonic_formula}
\begin{eqnarray}
\frac{d\Gamma(D\ra h\ell^+\nu)}{dq^2} & = & 
\frac{G^2_F}{24\pi^3} |f^{}_+(q^2)|^2 |V^{}_{cx}|^2 p^{*3}\,,
\label{eqn:semileptonic} 
\end{eqnarray}
where $h = K$ or $\pi$, 
$p^*$ is the magnitude of the $K$ or $\pi$ momentum in the 
$D$ rest frame, and $f^{}_+(q^2)$ is a form factor evaluated 
at $q^2=(P^{}_D - P^{}_h)^2 = (P^{}_\ell + P^{}_\nu)^2$.
If $h\!=\!K$, $V^{}_{cx}\!=\!V^{}_{cs}$, while if $h\!=\!\pi$, 
$V^{}_{cx}\!=\!V^{}_{cd}$. The form factor parameterizes the hadronic 
matrix element $\langle h|{\cal H}|D\rangle$ and is often modeled 
with a simple pole: $f_+(q^2) = f^{}_+(0)/(1-q^2/m^2_{\rm pole})$. One
thus fits the data at several values of $q^2$ to determine the 
normalization $f^{}_+(0)|V^{}_{cx}|$ and the parameter $m^{}_{\rm pole}$.

\begin{figure}[htb]
\begin{center}
\hbox{\hskip-0.10in
\includegraphics[width=6.3cm]{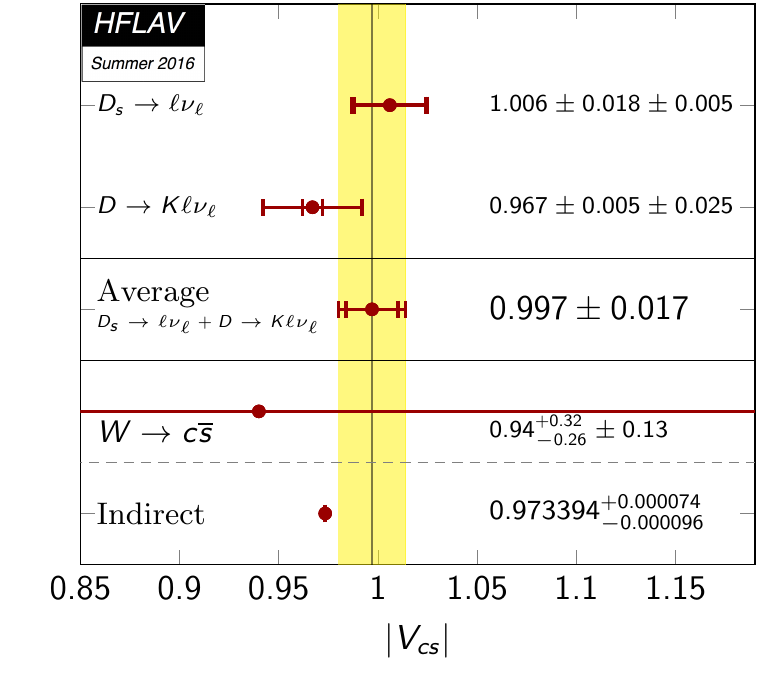}
\includegraphics[width=6.3cm]{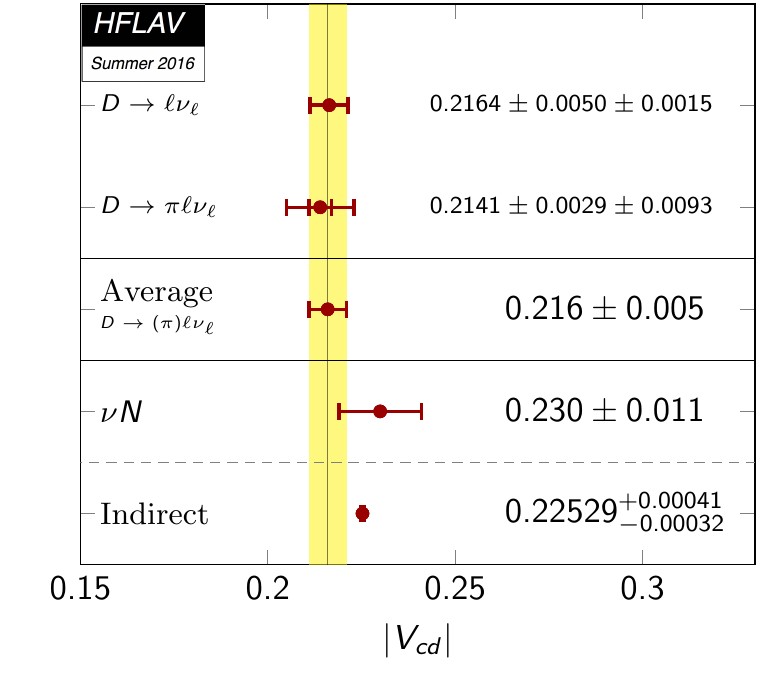}}
\end{center}
\vskip-0.30in
\caption{World average (WA) values of $|V^{}_{cs}|$ (left) and 
$|V^{}_{cd}|$ (right) as calculated by HFLAV using various 
methods~\cite{hflav:semileptonic}.}
\label{fig:hflav_compare}
\end{figure}

HFLAV has calculated WA values of $f^{}_+(0) |V^{}_{cx}|$ using relevant 
experimental measurements. The results are~\cite{hflav:semileptonic}
\begin{eqnarray}
f^K_+(0) |V^{}_{cs}| & = & 0.7226\,\pm 0.0022\,\pm 0.0026 \\
f^\pi_+(0) |V^{}_{cd}| & = & 0.1426\,\pm 0.0017\,\pm 0.0008\,,
\end{eqnarray}
where the first error is experimental and the second is from theory.
The Flavor Lattice Averaging Group~\cite{lattice:flag17} quotes results
$f^K_+(0) = 0.747\,\pm 0.019$ and $f^\pi_+(0) = 0.666\,\pm 0.029$ as
calculated by the HPQCD Collaboration~\cite{lattice:hpqcd11,lattice:hpqcd10}. 
Inserting these values gives 
\begin{eqnarray}
|V^{}_{cs}| & = & 0.967\,\pm 0.005 {\rm\ (exp.)}\,\pm 0.025 {\rm\ (theory)} 
\label{eqn:hflav_vcs} \\
|V^{}_{cd}| & = & 0.2141\,\pm 0.0029 {\rm\ (exp.)}\,\pm 0.0093 {\rm\ (theory)}\,.
\label{eqn:hflav_vcd}
\end{eqnarray}
These values have smaller experimental errors than those obtained from 
$D^+_{(s)}\ra\ell^+\nu$ decays, but the theory errors are larger. This reflects
the fact that experiments reconstruct much larger samples of semileptonic 
decays than purely leptonic decays, but
lattice QCD calculations of $f^{}_+(0)$ are less precise than
calculations of $f^{}_{D_{(s)}}$. A comparison of the different methods
made by HFLAV is shown in Fig.~\ref{fig:hflav_compare}. 
A recent calculation~\cite{lattice:riggio18} of the CKM matrix elements 
using lattice QCD results that account for the $q^2$ dependence of 
$f^{}_+$~\cite{lattice:etm17} gives $|V^{}_{cs}| = 0.970\,\pm 0.033$ 
and $|V^{}_{cd}| = 0.2341\,\pm 0.0074$. These values are consistent 
with results (\ref{eqn:hflav_vcs}) and~(\ref{eqn:hflav_vcd}).

\begin{center}
------------------
\end{center}

Belle has measured semileptonic $D^{*+}\ra D^0\pi^+$, 
$D^0\ra (K,\pi)^-\ell^+\nu$ decays using 282~fb$^{-1}$ 
of data~\cite{belle:semilept}. This analysis proceeds in four 
steps as done for the Belle $D^+_s\ra\ell^+\nu$ analysis:
\begin{enumerate}
\item a $D^{(*)0}$ or $D^{(*)+}$ decay is reconstructed on the 
tag-side of an event.
\item a fragmentation system $X^{}_{\rm frag}$ is constructed from 
remaining $\pi^\pm$ tracks, $K^\pm$ tracks (an even number), and 
$\pi^0$ candidates. The  missing mass
$M^{}_{\rm miss} = \sqrt{(P^{}_{e^+} + P^{}_{e^-} - P^{}_{\rm tag} - P^{}_{X_{\rm frag}})^2}$ 
is calculated, and a kinematic fit is performed subject to the constraint 
$M^{}_{\rm miss}=M(D^*)$. The resulting confidence level of the fit
is required to be $>\!0.1$\%, which corresponds to 
$M^{}_{\rm miss}$ being within $3.3\sigma$ of $M(D^*)$.
\item a low-momentum $\pi^+$ is selected from among remaining tracks, 
presumably originating from $D^{*+}\ra D^0\pi^+$, and the missing mass
$M'_{\rm miss}=
\sqrt{(P^{}_{e^+} + P^{}_{e^-} - P^{}_{\rm tag} - P^{}_{X_{\rm frag}}-P^{}_{\pi^+})^2}$ 
is calculated. A kinematic fit subject to the constraint
$M'_{\rm miss}=M(D^0)$ is performed, and the resulting confidence 
level is required to be $>\!0.1$\%. The $M'_{\rm miss}$ distribution 
is fitted to obtain an inclusive $D^0$ signal yield.
\item a $K^-$ or $\pi^-$ track, and also a $\mu^+$ or $e^+$ track, 
are required. No additional (signal candidate) tracks are allowed. 
The missing mass squared
$(P^{}_{e^+} + P^{}_{e^-} - P^{}_{\rm tag} - 
P^{}_{X_{\rm frag}}-P^{}_{\pi^+}-P^{}_{\ell^+} - P^{}_{(K,\pi)^-})^2$ is calculated.
For signal decays this quantity should equal $|m^{}_\nu|^2$, and thus it
is required to be $<\!0.05$~GeV$^2$/$c^4$.
\end{enumerate}
The signal yields are obtained after subtracting backgrounds. The results are
$2567 \pm 52\,{\rm (stat.)} \pm 26\,{\rm (syst.)}$ $D^0\ra K^-\ell^+\nu$ decays, 
and 
$232 \pm 17\,{\rm (stat.)} \pm 7\,{\rm (syst.)}$ $D^0\ra \pi^-\ell^+\nu$ decays. 

\begin{figure}[htb]
\begin{center}
\includegraphics[width=6.3cm]{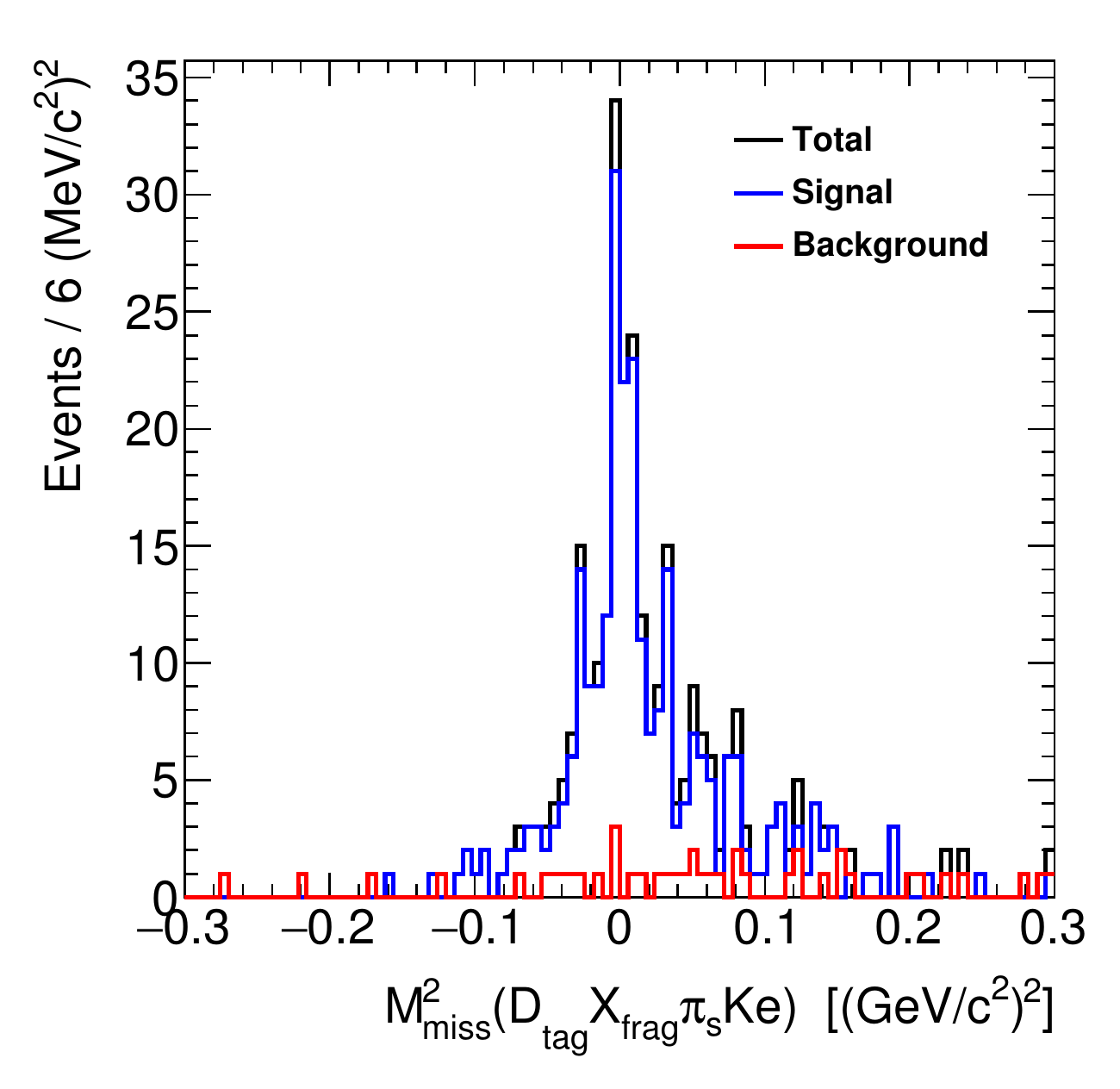}
\end{center}
\vskip-0.20in
\caption{
Missing mass squared
$(P^{}_{e^+} + P^{}_{e^-} - P^{}_{\rm tag} - P^{}_{X_{\rm frag}} - 
P^{}_{\pi^+} - P^{}_{K^-}- P^{}_{e^+})^2$
from a Belle II MC study of $D^{*+}\ra D^0\pi^+,\,D^0\ra K^- e^+\nu$ 
decays corresponding to 1~ab$^{-1}$ of data~\cite{belle2:ptep}. The 
peak at zero results from the undetected neutrino. For this study
only a single tag mode ($D^0\ra K^-\pi^+$) and a single fragmentation 
system ($X^{}_{\rm frag}\!=\!\pi^\pm$) were reconstructed.}
\label{fig:semi_belleII}
\end{figure}

This method can also be used at Belle~II. As the Belle measurement
was statistics- rather than systematics-limited, we simply scale the event 
yields obtained by Belle by the ratio of luminosities. The results are
455000 $D^0\ra K^-\ell^+\nu$ decays and 41100 $D^0\ra \pi^-\ell^+\nu$ decays 
in 50~ab$^{-1}$ of Belle~II data. An MC study of semileptonic decays in 
Belle~II~\cite{belle2:ptep} confirms that these analyses should have very 
low backgrounds and be statistics limited; see Fig.~\ref{fig:semi_belleII}.

\section{$D^0\ra \{ {\rm nothing}\}$ decays}

\begin{figure}[htb]
\begin{center}
\hbox{
\includegraphics[width=6.3cm]{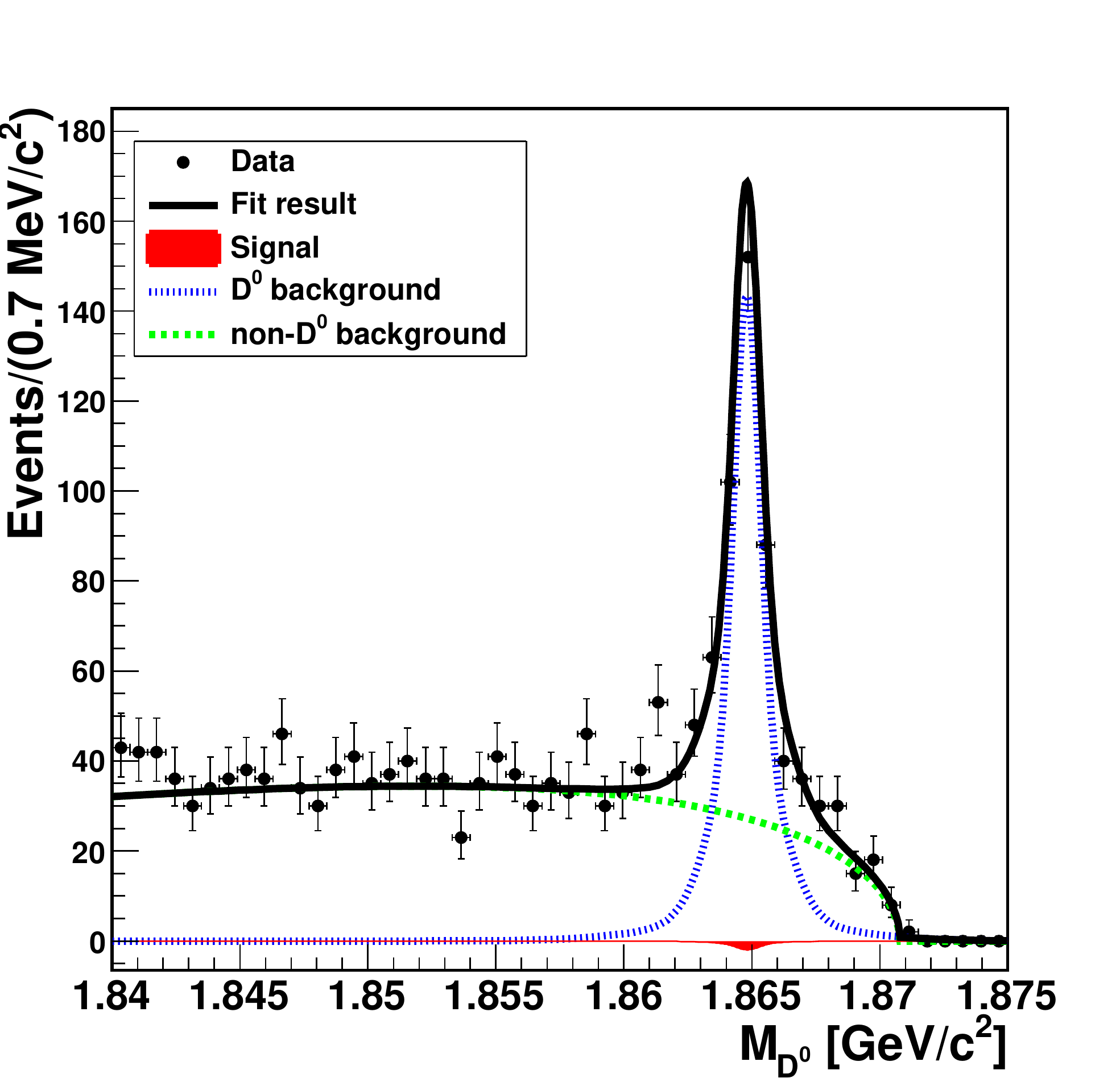}
\includegraphics[width=6.3cm]{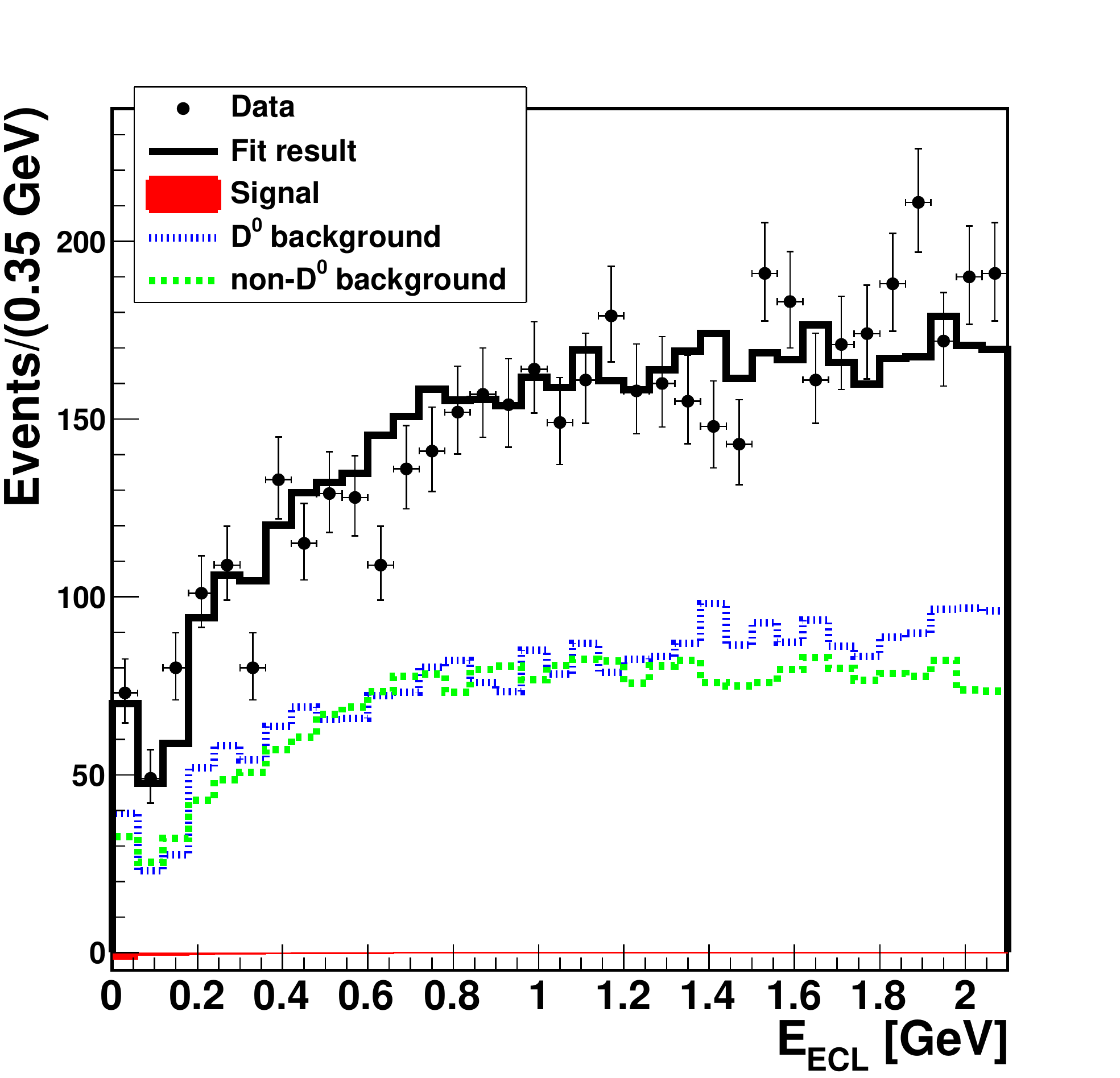}
}
\end{center}
\vskip-0.20in
\caption{
Final $D^{*+}\ra D^0\pi^+,\,D^0\ra\nu\bar{\nu}$ candidate sample from 
a Belle analysis of 924~fb$^{-1}$ of data~\cite{belle:nunu}.
Left: missing mass 
$\sqrt{(P^{}_{e^+} + P^{}_{e^-} - P^{}_{\rm tag} - P^{}_{X_{\rm frag}} - P^{}_{\pi^+})^2}$.
Right: excess energy deposited in the electromagnetic calorimeter (ECL),
i.e., ECL energy unassociated with any track. The $D^0\ra\nu\bar{\nu}$ 
signal yield (shown in red) is obtained by simultaneously fitting
both distributions.}
\label{fig:Dnunu_belle}
\end{figure}

In addition to measuring leptonic and semileptonic decays, Belle~II 
can search for the flavor-changing neutral current decay 
$D^0\ra\nu\bar{\nu}$, or more empirically, $D^0\ra \{ {\rm nothing}\}$. 
The SM rate is negligibly small ($1.1\times 10^{-30}$~\cite{Petrov:nunu}), 
and thus any 
evidence for this decay would indicate new physics. Belle searched
for this decay using 924~fb$^{-1}$ of data~\cite{belle:nunu}, and an 
analysis at Belle~II would proceed in a similar manner. As done for
the Belle analyses of leptonic and semileptonic decays, this analysis 
first reconstructs a tag-side $D^{(*)}$ decay. It then identifies a 
$\pi^+$ candidate originating from a signal-side $D^{*+}\ra D^0\pi^+$ 
decay; all remaining tracks are considered the fragmentation system 
$X^{}_{\rm frag}$. The missing mass 
$\sqrt{(P^{}_{e^+} + P^{}_{e^-} - P^{}_{\rm tag} - P^{}_{X_{\rm frag}})^2}$ 
is calculated and required to be near $M^2(D^*)$. 
The signal yield is calculated by simultaneously fitting 
two distributions: the ``$D^0$ missing mass''
$\sqrt{(P^{}_{e^+} + P^{}_{e^-} - P^{}_{\rm tag} - 
P^{}_{X_{\rm frag}}-P^{}_{\pi^+})^2}$, and the distribution of excess 
energy deposited in the electromagnetic calorimeter (ECL), i.e., 
energy clusters unassociated with any track.
These distributions are shown in Fig.~\ref{fig:Dnunu_belle}. 
No signal above background is observed, and an upper limit 
$B(D^0\ra\nu\bar{\nu})< 9.4\times 10^{-5}$ at 90\% C.L. is obtained. 
The size of the inclusive $D^0\ra\{{\rm nothing}\}$ sample 
is $694670\,^{+1490}_{-1560}$ events. Scaling this yield by the 
ratio of Belle and Belle~II luminosities gives 
$38\times 10^6$ inclusive $D^0$ decays in 50~ab$^{-1}$ 
of Belle~II data. Scaling the Belle single-event-sensitivity 
for $D^0\ra\{{\rm nothing}\}$ by a factor of $\sqrt{50/0.924}$ 
(the argument is the ratio of luminosities) implies 
a Belle~II upper limit of $1.3\times 10^{-5}$ at 90\% C.L.

\section{Summary}

Belle~II will measure leptonic $D^+_{(s)}\ra\ell^+\nu$ decays and 
semileptonic $D^0_{(s)}\ra (K,\pi)^-\ell^+\nu$ decays using methods
developed and refined at Belle and BaBar. In this paper we have
reviewed several Belle analyses of these decays and used the
results to estimate the sensitivity of Belle~II. 
As these measurements are dominated by statistical uncertainties, our 
estimates are based on scaling the Belle signal yields by the ratio of 
luminosities of Belle and Belle~II. 

The decays $D^+_s\ra\ell^+\nu$ and $D^+\ra\ell^+\nu$ 
constrain the products $f^{}_{D_s}|V^{}_{cs}|$ and $f^{}_{D}|V^{}_{cd}|$, 
respectively, and the decays
$D\ra K\ell^+\nu$ and $D\ra\pi\ell^+\nu$ constrain the products
$f^{K}_+(0)|V^{}_{cs}|$ and $f^{\pi}_+(0)|V^{}_{cd}|$, respectively. 
Taking decay constants $f^{}_{D_s}$ and $f^{}_{D}$ and form 
factor normalizations $f^{K}_+(0)$ and $f^{\pi}_+(0)$ from lattice 
QCD calculations, one can constrain CKM elements 
$|V^{}_{cs}|$ and $|V^{}_{cd}|$. In this manner one tests
CKM unitarity and the SM paradigm. Current results show
consistency with unitarity. As Belle~II plans to record 
50~ab$^{-1}$ of data, i.e., $\sim\!50$ times the sample size recorded 
by Belle, the resulting errors on $|V^{}_{cs}|,\,|V^{}_{cd}|$ should 
be reduced by a factor of $\sqrt{50}\approx 7$. 
Belle~II will also search for the flavor-changing neutral-current 
decay $D^0\ra\nu\bar{\nu}$. The full data set of Belle~II
should yield 7 times the sensitivity of Belle, and possibly 
much larger, depending on improvements in detector performance 
and reconstruction algorithms.

\begin{center}
------------------
\end{center}


We thank the workshop organizers for hosting a well-run meeting with 
excellent hospitality. We are grateful to Andreas Kronfeld for 
reviewing this paper and giving valuable input.


\begin{thebibliography}{99}
\bibitem{charge-conjugates}
Charge-conjugate modes are implicitly included throughout this paper.
\end{thebibliography}
\end{document}